\documentclass[aps,prd,reprint,floatfix,nofootinbib,nolongbibliography,superscriptaddress]{revtex4-2}

\usepackage[T1]{fontenc}
\usepackage[utf8]{inputenc}
\usepackage{graphicx}
\usepackage[usenames,dvipsnames]{xcolor}
\usepackage{mathtools}
\usepackage{amssymb}
\usepackage{lmodern}
\usepackage{siunitx}
\usepackage{slashed}
\usepackage[final]{microtype}
\usepackage{hyperref}

\hypersetup{%
	pdftitle = {Masses and decay constants of (axial-)vector mesons at finite chemical potential},
	pdfauthor = {Pascal J.~Gunkel, Christian S.~Fischer},
	bookmarksopen = true,
	colorlinks = true,
	linkcolor = blue!50!black,
	citecolor = blue!50!black,
	urlcolor = blue!50!black
}

\newcommand{\commutator}[2]{\left[#1,#2\right]}
\newcolumntype{d}[1]{D{.}{.}{#1}}

\newcommand*{\+}{\hspace*{.08335em}}

\newcommand*{\Nf}{N_{\textup{f}}}
\newcommand*{\Nc}{N_{\textup{c}}}
\newcommand*{\dd}{\textup{d}}
\newcommand*{\ZT}{Z_{\textup{T}}}
\newcommand*{\ZL}{Z_{\textup{L}}}

\DeclareMathOperator{\Tr}{Tr}

\DeclareSIUnit{\eV}{\electronvolt}

\begin{document}
	
	\title{Masses and decay constants of (axial-)vector mesons at finite chemical potential}
	
	\author{Pascal J.~Gunkel}
	\email{pascal.gunkel@physik.uni-giessen.de}
	\affiliation{%
		Institut f\"ur Theoretische Physik, %
		Justus-Liebig-Universit\"at Gie\ss{}en, %
		35392 Gie\ss{}en, %
		Germany%
	}
	
	\author{Christian S.~Fischer}
	\email{christian.fischer@theo.physik.uni-giessen.de}
	\affiliation{%
		Institut f\"ur Theoretische Physik, %
		Justus-Liebig-Universit\"at Gie\ss{}en, %
		35392 Gie\ss{}en, %
		Germany%
	}
	\affiliation{%
		Helmholtz Forschungsakademie Hessen f\"ur FAIR (HFHF),
		GSI Helmholtzzentrum \\
		f\"ur Schwerionenforschung, Campus Gie{\ss}en,
		35392 Gie{\ss}en, Germany
	}
	
	\begin{abstract}
		We update our previous results for (pseudo-)scalar mesons at zero temperature and finite quark chemical potential 
		and generalize the investigation to include (axial-)vector mesons. We determine bound-state properties such 
		as meson masses and decay constants up to chemical potentials far in the first-order coexistence region. To extract 
		the bound-states properties, we solve the Bethe-Salpeter equation and utilize Landau-gauge quark and gluon 
		propagators obtained from a coupled set of (truncated) Dyson-Schwinger equations with $\Nf=2+1$ dynamical 
		quark flavors at finite chemical potential and vanishing temperature. For multiple (pseudo-)scalar and (axial-)vector mesons, 
		we observe constant masses and decay constants for chemical potentials up to the coexistence region of the first-order 
		phase transition thus verifying explicitly the Silver-Blaze property of QCD. Inside the coexistence region the pion 
		becomes more massive and its decay constants decrease, whereas corresponding quantities for the (axial-)vector mesons 
		remain (almost) constant. 
	\end{abstract}
	
	\maketitle
	
	\section{\label{sec:intro}Introduction}
	
	In the analysis of experimental heavy-ion collisions, electromagnetic radiation from the hot and dense
	fireball plays a pivotal role. Once the real or virtual photon is produced in the reaction, it escapes
	the medium almost undistorted and can therefore serve as a probe for the state of matter in the early
	stages of the collision. Due to their quantum numbers, vector mesons couple to the electromagnetic 
	current and therefore in particular the light ones, $\rho$, $\omega$, and $\phi$ are expected to contribute
	substantially to the observed dilepton spectrum. The study of the in-medium properties of vector mesons
	has thus received considerable attention, see, e.g., Refs.~\cite{Rapp:1999ej,Leupold:2009kz,Friman:2011zz} 
	for reviews. 
	
	The region of the QCD phase diagram with low temperatures and large densities is the realm of cold 
	nuclear matter. The properties of vector mesons, in particular their spectral functions have been
	studied in a range of approaches with focus on the medium effects of their pion cloud as 
	well as medium effects due to the coupling of the $\rho$ meson to nucleons via resonance 
	excitations \cite{Rapp:1999ej,Leupold:2009kz,Friman:2011zz}. A very recent study in this direction takes into 
	account quantum fluctuations via the functional renormalization group approach to a low-energy 
	effective theory \cite{Jung:2016yxl,Jung:2019nnr}. 
	
	Less studied is the direct impact of non-vanishing chemical potential on the quark and gluon structure 
	of the $\rho$, its chiral partner $a_1$ and the $\phi$ meson and the resulting changes in their mass 
	and decay constants. This is the topic of this
	work. Based on previous results for the chemical-potential dependence of the quark propagator and the 
	resulting behavior of the masses and decay constants of pseudo-scalar and scalar mesons at finite
	chemical potential \cite{Gunkel:2019xnh} we improve the approach and generalize it to also accommodate
	for vector and axial-vector mesons. Working at zero temperature, our study is complementary to a recent study of the thermal
	properties of vector mesons at and around the crossover at finite temperature \cite{Chen:2020afc}.
	
	One of the most interesting questions associated with the zero temperature, finite chemical potential
	axis of the QCD phase diagram is the Silver-Blaze property of QCD: For baryon chemical potentials
	smaller than the mass of the nucleon minus its binding energy in nuclear matter, there can be no excitations
	from the QCD vacuum and therefore all observables have to retain their vacuum values. This can be shown 
	analytically for the case of finite isospin chemical potential, but is also extremely plausible for the 
	case of finite baryon chemical potential \cite{Cohen:1991nk,Cohen:2004qp} and has been demonstrated for 
	heavy quark masses  in the lattice formulation of  Ref.~\cite{Fromm:2012eb}. How this works in the case of
	pseudoscalar and scalar mesons has been studied in Ref.~\cite{Gunkel:2019xnh}. Here, we will see that
	a similar mechanism is in place for the vector and axial-vector mesons.  
	
	The paper is organized as follows: In Sec.~\ref{sec:bse_formalism} we discuss the truncation of
	Dyson-Schwinger and Bethe-Salpeter equations that we use in our study. More details can be found in 
	Ref.~\cite{Gunkel:2019xnh}. In Sec.~\ref{sec:results} we present our updated results for the 
	chemical-potential dependence of masses and decay constants of pseudo-scalar and scalar mesons as well
	as results for the vector and axial-vector mesons. We conclude in Sec.~\ref{sec:summary}.

	\section{\label{sec:bse_formalism}Bethe-Salpeter formalism}
	
	The homogeneous Bethe-Salpeter equation (BSE) for (pseudo-)scalar and (axial-)vector mesons in ladder truncation 
	is given by
	\begin{align}
	\Gamma_{\textup{x},f}^{(\mu)}(p,P)=-(Z_2^f)^2\+g^2\+C_{\textup{F}}\int_{q}&\,\gamma_{\mu}\+S_{f}(q_{+})\+\Gamma_{\textup{x},f}^{(\mu)}(q,P) \nonumber \\
	&\times S_{f}(q_{-})\+\gamma_{\nu}\Gamma(k^2)\+D_{\mu\nu}(k)
	\label{eq:homogeneous_Bethe_salpeter_equation}
	\end{align}
	with the shorthand $\int_{q} \equiv \int \dd^{4}q \+ / \+ (2\pi)^{4}$ and the strong coupling constant
	$g^2=4\pi\alpha_\textup{s}$. 
	The Casimir $C_\textup{F}= (\Nc^{2}-1) \+ / \+ (2\Nc)$ results from the color trace with $\Nc=3$. $Z_2^f$ 
	represents the quark wave function renormalization constant of the quark flavour 
	$f\in\{\textup{u},\textup{d},\textup{s}\}$. The relative momenta of the meson ($p$ and $q$) entail 
	the quark chemical potential\footnote{In this work, we use a vanishing isospin $\mu_\textup{I}$ and 
		strangeness $\mu_\textup{S}$ chemical potential implying
		$\mu_\textup{q}^\textup{u}=\mu_\textup{q}^\textup{d}=\mu_\textup{q}^\ell$ and $\mu_\textup{q}^\textup{s}=\mu_\textup{q}^\ell$. 
		Furthermore, we often express the chemical potential by the baryon chemical potential $\mu_\textup{B}=3\mu_\textup{q}^\ell$.}
	$\mu_{\textup{q}}^{f}$ and are given by $p=(\vec{p},\tilde{p}_4)$ with $\tilde{p}_4=p_{4}+i\mu_{\textup{q}}^{f}$. 
	We consider the meson to be in its rest frame, i.e., $P=(\vec{0},im_\textup{x})$ with the time-like total 
	momentum $P$ and the mass $m_\textup{x}$ of the meson. The index x thereby describes the meson type.
	We use the momentum routings $k=p-q$ and $q_\pm=q\pm\eta_\pm P$ for the gluon and the quark momenta, respectively. 
	The momentum-partitioning parameters $0\leq\eta_\pm\leq1$ can be varied within the boundary condition $\eta_++\eta_-=1$.
	In the vacuum, Poincar$\acute{\textup{e}}$ covariance implies the independence of all observables on the choice of 
	$\eta_\pm$. Numerically, this is satisfied on the permille level. At finite chemical potential and for pseudoscalar mesons 
	we explicitly verified that this invariance holds for chemical potentials up and into the coexistence region. For larger 
	chemical potentials and for heavier mesons $\eta_+$ needs to be adapted such that the integration in the BSE avoids 
	the complex plane singularities in the quark propagators.
	 
	The homogeneous BSE depends on the dressed quark and gluon propagators $S_f$ and $D_{\mu\nu}$ as well as the dressed quark-gluon vertex
	with dressing function $\Gamma(k^2)$. A detailed discussion of our truncation for the vertex as well as explicit
	expressions for $\Gamma(k^2)$ and the corresponding choice of the coupling $\alpha_\textup{s}$ can be found in Ref.~\cite{Gunkel:2019xnh}.
	For finite quark 
	chemical potential $\mu_{\textup{q}}^{f}$ and vanishing temperature, the Landau-gauge propagators can be written as
	\begin{align}
	S^{-1}_f(p)=&\;i\vec{\slashed{p}}\+A_f(p)+i\tilde{p}_{4}\gamma_{4}\+C_f(p)+B_f(p)\,, \\
	D_{\mu\nu}(k)=&\;\mathcal{P}_{\mu\nu}^{\textup{T}}(k) \+ \frac{\ZT(k)}{k^{2}}+\mathcal{P}_{\mu\nu}^{\textup{L}}(k)\+\frac{\ZL(k)}{k^{2}}.
	\label{eq:dressed_quark_gluon_propagator_ansatz_landau_gauge}
	\end{align}
	Here, the four-dimensional transverse projector is split into a part $\mathcal{P}_{\mu\nu}^{\textup{T}}(k)$ 
	transverse to the assigned direction $v=(\vec{0},1)$ of the medium and a corresponding longitudinal part $\mathcal{P}_{\mu\nu}^{\textup{L}}(k)$.
    The associated dressing functions of the gluon split into the transverse (or magnetic) part $\ZT$ and the longitudinal
    (or electric) part $\ZL$.    
	The gluon and quark ($A_f$, $B_f$, $C_f$) dressing functions encode the non-trivial momentum 
	dependence of the propagators. For vanishing chemical potential the vector dressing functions $A_f$ and $C_f$ 
	as well as the gluon dressing functions degenerate. For finite chemical potential this is in general no
	longer the case. 
	
	The quark and gluon propagators are calculated from a coupled set of truncated Dyson-Schwinger equations. 
	In the corresponding truncation, we use quenched lattice data for the gluon as input and unquench it explicitly 
	by including the back-reaction of the quark onto the gluon. Furthermore, we use an ansatz for the quark-gluon 
	vertex. Details for this well-studied truncation can be found in Ref.~\cite{Eichmann:2015kfa} and the review
	Ref.~\cite{Fischer:2018sdj}. In previous works this truncation was used for the quenched case \cite{Fischer:2009wc,Fischer:2010fx} as well as for different numbers of quark flavors \cite{Fischer:2011pk,Fischer:2011mz,Fischer:2012vc,Fischer:2014ata}. Also supercolorconductivity has been studied using this truncation \cite{Muller:2013pya,Muller:2016fdr}. Two further approximations were made in the preceding work of Ref.~\cite{Gunkel:2019xnh}: (i) the chemical-potential dependence of the gluon is neglected and (ii) a slightly modified quark-gluon vertex ansatz in the quark DSE is used. 
	We build upon this work and adopt these approximations.
	
	In the vacuum BSE we use the same tensor-structure decomposition for pseudoscalar ($\textup{x}=\textup{P}$) and scalar ($\textup{x}=\textup{S}$) mesons as 
	detailed in Ref.~\cite{Gunkel:2019xnh}. In the medium we extend the Bethe-Salpeter amplitude (BSA) to
	\begin{align}
	\label{eq:P_S_BSA_finite_mu}
	\Gamma_{\textup{P}}(p,P)=&\;\gamma_{5}\left\{E_{\textup{P}}(p,P)-i\vec{\slashed{p}}\,P\cdotp p\,G_\textup{P}^\textup{s}(p,P)\;+\right. \nonumber \\
	&\left.\phantom{\;\gamma_{5}\{}-i\gamma_{4} \+ I_{\textup{P}}(p,P)\right\}, \\[0.5em]
	\Gamma_{\textup{S}}(p,P)=&\,\,\;\openone_\textup{D}\left\{E_{\textup{S}}(p,P)-i\vec{\slashed{p}}\,G_\textup{S}^\textup{s}(p,P)-i\gamma_{4} \+ I_{\textup{S}}(p,P)\right\}\,,
	\end{align}
	including the additional structure $G_\textup{x}^\textup{s}$ as compared to Ref.~\cite{Gunkel:2019xnh}. Results from the vacuum suggest that
	this addition will not change the meson masses by much, but may be relevant for the decay constant \cite{Maris:1997hd}. We will see later, that
	this is indeed the case.
	The flavor dependence of the amplitude is suppressed in our notation. 
	For the vector ($\textup{x}=\textup{V}$) and axial-vector ($\textup{x}=\textup{A}$) in vacuum we work with the tensor decomposition detailed in previous works (see, e.g., Refs.~\cite{Smith:1969,Fischer:2005en,Williams:2009wx}):
	\begin{align}
	\label{eq:V_AV_BSA_vacuum}
	\Gamma_\textup{V}^\mu(p,P)=&\;i\gamma_\top^\mu F_{1\textup{V}}(p,P)+\gamma_\top^\mu\slashed{P} F_{2\textup{V}}(p,P) \nonumber \\
	&\;+(p_\top^\mu\openone_\textup{D}-\gamma_\top^\mu\slashed{p})\,P\cdotp p\, F_{3\textup{V}}(p,P) \nonumber \\
	&\;+(i\gamma_\top^\mu\commutator{\slashed{P}}{\slashed{p}}+2ip_\top^\mu\slashed{P}) F_{4\textup{V}}(p,P) \nonumber \\
	&\;+p_\top^\mu\openone_\textup{D} F_{5\textup{V}}(p,P)+ip_\top^\mu\slashed{P}\,P\cdotp p\, F_{6\textup{V}}(p,P) \nonumber \\
	&\;-ip_\top^\mu\slashed{p} F_{7\textup{V}}(p,P)+p_\top^\mu\commutator{\slashed{P}}{\slashed{p}} F_{8\textup{V}}(p,P), \nonumber \\[0.5em]
	\Gamma_\textup{A}^\mu(p,P)=&\;\gamma_5\left\{i\gamma_\top^\mu F_{1\textup{A}}(p,P)+\gamma_\top^\mu\slashed{P} \,P\cdotp pF_{2\textup{A}}(p,P) \right. \nonumber \\
	&\;+(p_\top^\mu\openone_\textup{D}-\gamma_\top^\mu\slashed{p})\, F_{3\textup{A}}(p,P) \nonumber \\
	&\;+(i\gamma_\top^\mu\commutator{\slashed{P}}{\slashed{p}}+2ip_\top^\mu\slashed{P}) F_{4\textup{A}}(p,P) \nonumber \\
	&\;+p_\top^\mu\openone_\textup{D} \,P\cdotp pF_{5\textup{A}}(p,P)+ip_\top^\mu\slashed{P}\,P\cdotp p\, F_{6\textup{A}}(p,P) \nonumber \\
	&\;\left.-ip_\top^\mu\slashed{p} F_{7\textup{A}}(p,P)+p_\top^\mu\commutator{\slashed{P}}{\slashed{p}} \,P\cdotp pF_{8\textup{A}}(p,P)\right\}.
	\end{align}
	The tensor decomposition is constructed such that the on-shell (axial-)vector meson is transverse to its total momentum $P$. 
	The subscript $\top$ indicates transversality of $w\in\{\gamma,p\}$ w.r.t.~the total momentum, i.e., $w_\top^\mu=T_{\mu\nu}(P)w^\nu$.
	
	In medium, the number of independent transverse tensor structures increases from eight to 24 thus inducing considerable numerical costs.
	Therefore we only consider two BSA components $F_{1\textup{x}}$ and $F_{4\textup{x}}$ for the qualitative study of this work. While $F_{1\textup{x}}$ is the dominant BSA component of the vector meson, $F_{4\textup{x}}$ is the correspondingly
	dominant one for axial-vector mesons. In medium the BSA components splits 
	up into a spatial $F_{i\textup{sV}}$ and a temporal $F_{i\textup{tV}}$ component implying two separate uncoupled BSEs 
	to solve. The employed tensor decomposition in medium is
	\begin{align}
	\label{eq:V_AV_BSA_finite_mu}
	\Gamma_\textup{V}^\mu(p,P)=&\;i\gamma_{\top\textup{T}}^\mu F_{1\textup{sV}}(p,P) \nonumber \\
	&+(i\gamma_{\top\textup{T}}^\mu\commutator{\slashed{P}}{\slashed{p}}+2ip_{\top\textup{T}}^\mu\slashed{P})F_{4\textup{sV}}(p,P) \nonumber \\
	&+i\gamma_{\top\textup{L}}^\mu F_{1\textup{tV}}(p,P)+2ip_{\top\textup{L}}^\mu P_4\gamma_4 F_{4\textup{tV}}(p,P), \nonumber \\[0.5em]
	\Gamma_\textup{A}^\mu(p,P)=&\;\gamma_5\left\{i\gamma_{\top\textup{T}}^\mu F_{1\textup{sA}}(p,P)\right. \nonumber \\
	&+(i\gamma_{\top\textup{T}}^\mu\commutator{\slashed{P}}{\slashed{p}}+2ip_{\top\textup{T}}^\mu\slashed{P})F_{4\textup{sA}}(p,P) \nonumber \\
	&+\left.i\gamma_{\top\textup{L}}^\mu F_{1\textup{tA}}(p,P)+2ip_{\top\textup{L}}^\mu P_4\gamma_4 F_{4\textup{tA}}(p,P)\right\}
	\end{align}
	with $w_{\top(\textup{T}/\textup{L})}^\mu=\mathcal{P}_{\mu\nu}^{\textup{T}/\textup{L}}(P)w^\nu$ and $w\in\{\gamma,p\}$. So far the $\rho$ meson at finite temperature was investigated using 
	only the dominant BSA component $F_{1\textup{V}}$ with an effective interaction (see, e.g., Refs.~\cite{Maris:2001rq,Maris:1997eg,Maris:1999nt}).
	
	All amplitudes of the (pseudo-)scalar and (axial-)vector mesons are normalized using the Nakanishi method \cite{Nakanishi:1965zza} and serve as
	input into the calculation of the pseudo-scalar and (axial-)vector meson decay constants $f_{\textup{x}}$. In vacuum these are given by  
	\begin{align}
	\label{eq:decay_constants}
	f_\textup{x}=\frac{\Nc}{im_\textup{x}}\int_q\Tr_\textup{D}\left\{j_{\textup{x},f}^{(\mu)}(P)S_f(q_+)\hat{\Gamma}_{\textup{x},f}^{(\mu)}(q,P)S_f(q_-)\right\},
	\end{align}
	where $\hat{\Gamma}_{\textup{x},f}^{(\mu)}$ represents the normalized BSA and the current $j_{\textup{x},f}^{(\mu)}(P)$ is defined by
	\begin{align}
	\label{eq:decay_constants_currents}
	j_{\textup{x},f}^{(\mu)}(P)=
	\begin{cases}
	Z_2^f\gamma_5\hat{\slashed{P}} & \textup{for x}=\textup{P} \\
	Z_2^f\gamma_\top^\mu\tfrac{1}{3} & \textup{for x}=\textup{V} \\
	Z_2^f\gamma_5\gamma_\top^\mu\tfrac{1}{3} & \textup{for x}=\textup{A}
	\end{cases}.
	\end{align}
	Equation \eqref{eq:decay_constants} is evaluated for the on-shell momentum $P^2=-m_\textup{\textup{x}}^2$ and is exact if the dressed 
	quark propagators and the meson BSA are exact. The factor $1/3$ in the (axial-)vector case has to be included because 
	of the summation over the polarizations.
	
	In medium the decay constant of the pseudo-scalar meson splits into two parts \cite{Son:2001ff,Son:2002ci} as discussed in our previous 
	work \cite{Gunkel:2019xnh}. One arrives at
	\begin{align}
	\label{eq:pseudoscalr_decay_constants_medium}
	f_\textup{P}^{\textup{s}/\textup{t}}=\Nc\int_q\Tr_\textup{D}&\left\{j_{\textup{P},f}^{\textup{s}/\textup{t}}(P) S_f(q_+)\hat{\Gamma}_{\textup{P},f}^\mu(q,P)S_f(q_-)\right\},
	\end{align}
	with the corresponding current
	\begin{align}
	\label{eq:pseudoscalar_decay_constants_current_medium}
	j_{\textup{P},f}^{\textup{s}/\textup{t}}(P)=
	\begin{cases}
	Z_2^f\gamma_5\vec{\slashed{P}}/\vec{P}^2 & \textup{for s} \\
	Z_2^f\gamma_5\gamma_4/P_4 & \textup{for t}
	\end{cases}.
	\end{align}
	For the (axial-)vector meson in medium we can equally define a spatial and temporal decay constant belonging to the 
	spatial and temporal BSA, respectively:
	\begin{align}
	\label{eq:vector_decay_constants_medium}
	f_\textup{x}^{\textup{s}/\textup{t}}=\frac{\Nc}{im_x}\int_q\Tr_\textup{D}&\left\{j_{\textup{x},f}^{\textup{s}/\textup{t},\mu}(P) S_f(q_+)\hat{\Gamma}_{\textup{x},f}^\mu(q,P)S_f(q_-)\right\}
	\end{align}
	In this equation the current for vector mesons is defined as
	\begin{align}
	\label{eq:vector_decay_constants_current_medium}
	j_{\textup{V},f}^{\textup{s}/\textup{t},\mu}(P)=
	\begin{cases}
	Z_2^f\gamma_{\top\textup{T}}^\mu\tfrac{1}{2} & \textup{for s} \\
	Z_2^f\gamma_{\top\textup{L}}^\mu & \textup{for t}
	\end{cases}.
	\end{align}
	In case of axial-vector mesons a $\gamma_5$ factor has to be included in the the current.

	\section{\label{sec:results}Results}
	
	%
	\begin{figure}[t]
		\centering
		\includegraphics[width=0.47\textwidth]{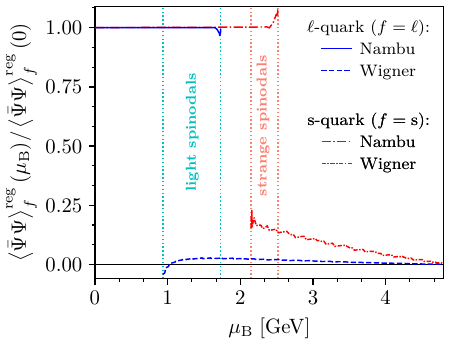}%
		\caption{\label{fig:light_strange_quark_condensate}%
			Vacuum normalized and regularized light (blue) and strange (red) quark condensate for the chirally-broken Nambu (solid and dashed lines) and chirally-restored Wigner (differently dashed dotted lines) solution plotted against the baryon chemical potential $\mu_\textup{B}$. The boundaries for the appearance/ disappearance of the Nambu and Wigner solution are denoted by vertical dotted lines in the corresponding color of the flavor and called light and strange spinodals.}
	\end{figure}
	\begin{figure*}[t]
		\centering
		\includegraphics[width=0.45\textwidth]{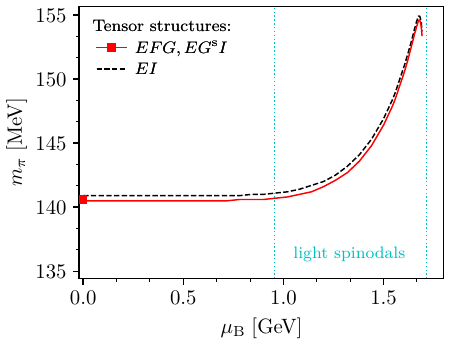}%
		\hfill
		\includegraphics[width=0.45\textwidth]{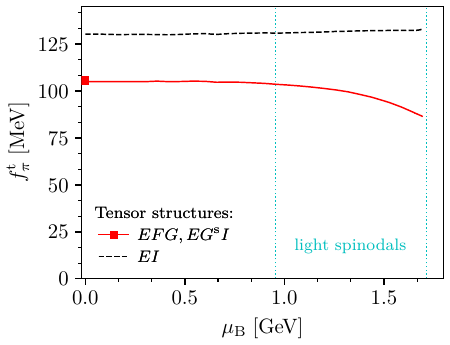}%
		\caption{\label{fig:pion_properties_finite_mu}%
			Pion mass (left) and temporal pion decay constant (right) against the baryon chemical potential for different combinations of tensor structures used in the BSE calculation. The colored symbols represent the corresponding vacuum results. All results are obtained with the chirally-broken Nambu solution. The vertical lines indicate the boundaries of the light quark coexistence region.}
	\end{figure*}
	In this section, we discuss our results for the masses and decay constants of light and strange (pseudo)-scalar and (axial-)vector mesons
	at non-vanishing chemical potential. In the discussion, we denote by 'coexistence region' the region of $\mu_B$ where both, the chirally-broken 
	Nambu solution and the chirally symmetric Wigner solution are available and attractive in the iteration process. The 'spinodal points' are the 
	end points of this region. For smaller chemical potentials the Wigner solution is still present (see e.g. \cite{Fischer:2008sp}) but is not 
	iteratively attractive. For larger chemical potentials the Nambu solution ceases to exist. 
	
	 Note that in principle it is possible to determine the thermodynamic potential in our approach and therefore determine the boundary
		of first order transition within the coexistence region. However, for the coupled system of DSEs that we use in this work this is a
		non-trivial numerical task that requires considerable additional effort. Since thermodynamics is not the main issues of this work 
		we postponed this task to a future work.
	
	All our results are calculated using the Nambu 
	solution of the quark DSE. The Wigner solution features poles close to the time-like real momentum axis at very low masses, posing technical 
	problems in the Bethe-Salpeter equation which are beyond current solution techniques. 
	
	In Ref.~\cite{Gunkel:2019xnh} we studied the areas of stability of the Nambu and Wigner solution for the light quark at finite (light) chemical potential and 
	vanishing temperature and located the coexistence region where both solutions exist and are stable. Here, in contrast to the previous work we additionally 
	use a non-vanishing strange-quark chemical potential and investigate the simplest case $\mu_\textup{q}^\textup{u}=\mu_\textup{q}^\textup{d}=\mu_\textup{q}^\textup{s}$. In the fully back-coupled system, the first 
	order transition of the up/down quark sector at some critical chemical potential then necessarily introduces non-analytic changes also in the strange quark 
	propagator (see e.g. \cite{Otto:2019zjy} for an explicit calculation of this effect). However, is is not clear whether the loss in interaction strength due 
	to almost massless (and screening) up/down quarks is sufficient to reduce the strange quark immediately to its Wigner solution. Instead it might be that 
	within some region of chemical potential the strange quark still feels dynamical chiral symmetry breaking (however with reduced strength) whereas the up/down 
	sector is already in the Wigner mode. Then at even larger chemical potential a second first order transition in the strange quark sector would occur. Whether 
	this scenario is realistic is an open question that remains to be studied.
	
	In this work, however, we neglect the chemical-potential dependency of the gluon by always using the (unquenched) gluon propagator from the vacuum.\footnote{In Ref.~\cite{Contant:2019lwf} the influence of this approximation was studied for the case of two color QCD.} Thus, any 
	changes in the light and strange quark sector induced by chemical potential are not back-coupled to the respective other sector. Therefore, we do expect to find 
	two different coexistence areas for the light and strange quark. Indeed, this can be seen in Fig.~\ref{fig:light_strange_quark_condensate}, where we show our results 
	for the vacuum normalized and regularized\footnote{We regularize the quark condensate by subtracting the quark condensate at very high chemical potential, 
		where the dynamical part is expected to vanish:
		$\left\langle\bar{\Psi}\Psi\right\rangle_f^\textup{reg}(\mu_\textup{B})
		=\left\langle\bar{\Psi}\Psi\right\rangle_f(\mu_\textup{B})-\left\langle\bar{\Psi}\Psi\right\rangle_f(\infty)$.} 
	light $\left\langle\bar{\Psi}\Psi\right\rangle_\ell$ and strange $\left\langle\bar{\Psi}\Psi\right\rangle_\textup{s}$ quark condensate plotted against the 
	baryon chemical potential $\mu_\textup{B} = 3\mu_\textup{q}^\ell=3\mu_\textup{q}^\textup{s}$. We also plot the boundaries of the coexistence regions to guide the eye. We find these at:
	\begin{align}
	\begin{array}{cccc}
	&\textup{Wigner:}      & \textup{Nambu:}      &                   \\[0.3em]
	\mu_B=  &\SI{0.936}{\giga\eV}, & \SI{1.730}{\giga\eV} & (\textup{light}), \\[0.4em]
	\mu_B=  &\SI{2.149}{\giga\eV}, & \SI{2.516}{\giga\eV} & (\textup{strange})
	\end{array}
	\end{align}
	Slight changes as compared to Ref.~\cite{Gunkel:2019xnh} are due to improved numerics.
	
	\subsection{\label{sec:meson_properties_mu_dependence}Chemical-potential dependence of the meson properties}
	
	%
	\begin{figure*}[t]
		\centering
		\includegraphics[width=0.50\textwidth]{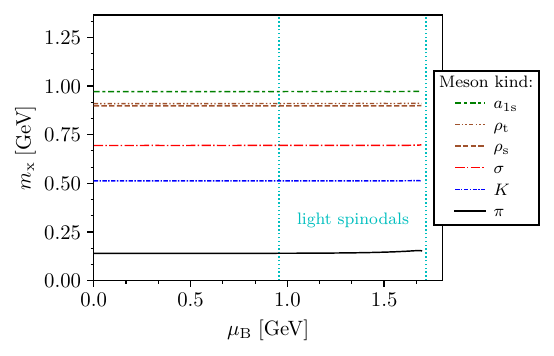}%
		\hfill
		\includegraphics[width=0.50\textwidth]{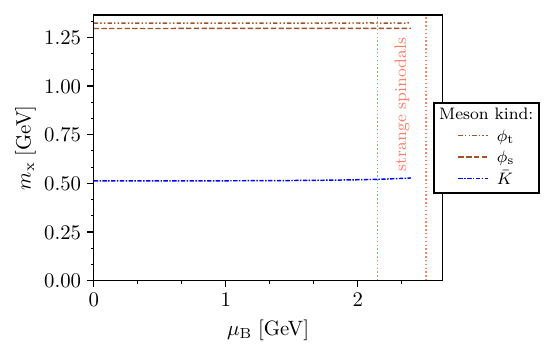}\\
		\includegraphics[width=0.50\textwidth]{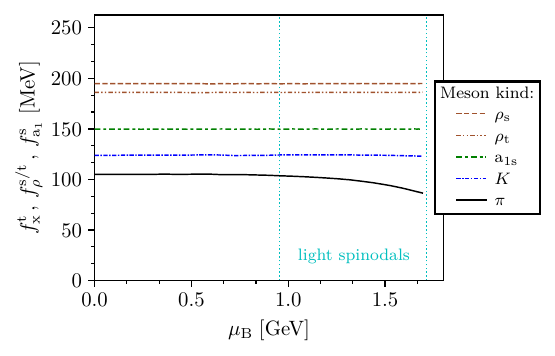}%
		\hfill
		\includegraphics[width=0.50\textwidth]{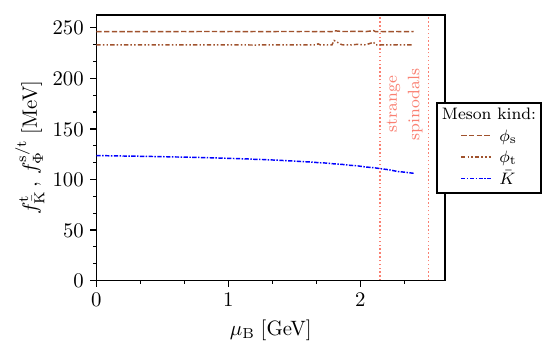}\\
		\caption{\label{fig:meson_properties_finite_mu}%
			Masses (upper) and decay constants (lower) for different light quark (left) and strange quark (right) mesons plotted against the baryon chemical potential for the most elaborated tensor structure combination of the BSE in medium. The results are calculated using the chirally-broken Nambu solution.}
	\end{figure*}

	In Fig.~\ref{fig:pion_properties_finite_mu} we display the pion mass and temporal decay constant in vacuum and at finite chemical potential for two 
	different levels of approximation of the Bethe-Salpeter amplitude (BSA). We show results from the most elaborated BSA truncation of Ref.~\cite{Gunkel:2019xnh} 
	(dashed line, black) and the improved truncation used in this work (solid line, red). A corresponding colored symbol displays the results of the vacuum calculation. For our improved truncation this limit is smooth and well defined, whereas it is ambiguous	for the truncation used in Ref.~\cite{Gunkel:2019xnh}.
	Therefore no symbol is shown for this case.  It furthermore turns out that the addition of $G^\textup{s}$ only has a small effect for the pion mass, 
	but has a significant quantitative effect of the order of 20 percent on the temporal decay constant, as anticipated above.
	
	Within the numerical precision the pion mass and temporal decay constant remain constant up to a baryon chemical potential equal to the mass of a nucleon $m_\textup{N}=\SI{0.928}{\giga\eV}$ in medium. In this region, the mass and decay constant deviate by less than $\SI{0.25}{\%}$ and $\SI{2}{\%}$ from their 
	vacuum values. This is constant within numerical accuracy. 
	Consequently, we can state that the pion properties fulfill the Silver-Blaze property. Until the end of the coexistence region the pion mass 
	increases up to $\SI{14}{\%}$ compared to the vacuum value. For the temporal pion decay constant from our improved truncation scheme we find a
	$\SI{20}{\%}$ decrease compared to the vacuum value.
	
	In Fig.~\ref{fig:meson_properties_finite_mu} we show the meson properties for multiple light and strange quark mesons at finite chemical potential 
	in our improved truncation scheme. The $K$ and $\bar{K}$ meson behave similarly as the $\pi$ meson. But while the observed decrease of the decay 
	constant is less pronounced for the $K$ meson, the contrary is true for the $\bar{K}$ meson. Overall the kaon masses increase by less than \SI{3}{\%}
	while the decay constant decrease by \SI{15}{\%}. 
	For the $\sigma$ meson there is no significant qualitative or quantitative difference between the different levels of approximation of the BSA
	as used in Ref.~\cite{Gunkel:2019xnh} and here (therefore we only display the updated result). The mass of the $\sigma$-meson and the longitudinal 
	and transversal $\rho$ and $\phi$ properties remain perfectly constant until the end of the corresponding coexistence area with a maximal deviation 
	of less than $\SI{0.5}{\%}$. Again we note that the Silver-Blaze property is very well satisfied.  

	\begin{figure*}[t]
		\centering
		\includegraphics[width=0.49\textwidth]{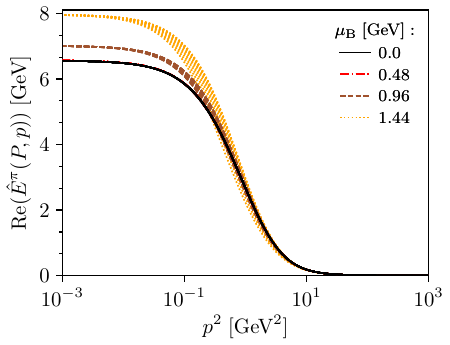}%
		\hfill
		\includegraphics[width=0.49\textwidth]{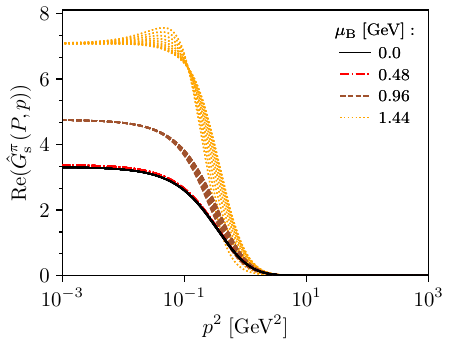}\\
		\centering
		\includegraphics[width=0.49\textwidth]{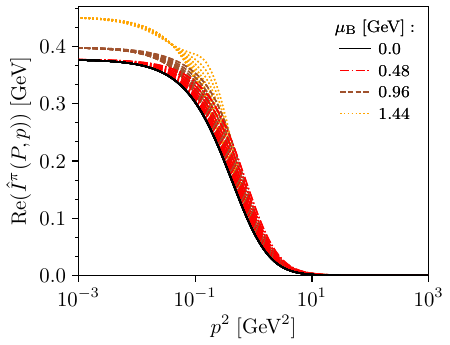}\\
		\centering
		\includegraphics[width=0.49\textwidth]{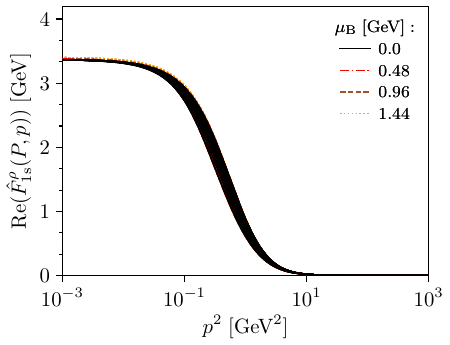}%
		\hfill
		\includegraphics[width=0.49\textwidth]{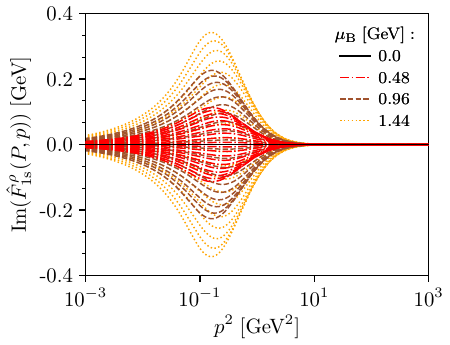}%
		\caption{\label{fig:pion_bsa_finite_mu}%
			Upper panel: Real part of the first and second normalized on-shell medium $\pi$-BSA component $\hat{E}^\pi$ (left) and $\hat{G}_\textup{s}^\pi$ (right) ploted against the relative momentum $p^2$ between the quark and the antiquark for various baryon chemical potentials $\mu_\textup{B}$ far into the coexistence region. The spread of the amplitude results from the dependence on the angle between $P$ and $p$. The results are calculated for the chirally-broken Nambu solution only.\\ 
			Middle panel: Real part of the third normalized on-shell medium $\pi$-BSA component $\hat{I}^\pi$.\\
			Lower panel: Real (left) and imaginary (right) part of the leading tensor component $\hat{F}^\rho_{1s}$ of the rho meson. For the real part, the chemical potential dependence of the BSA is very weak, whereas much stronger variations are visible in the imaginary part.}
	\end{figure*}

	The masses and decay constant of the temporal and spatial projections of the $\rho$ and $\phi$ meson have different values and we find that only the 
	spatial quantities have a smooth vacuum limit. The reason is that the temporal projections of the BSA at finite chemical potential receive contributions 
	from several vacuum BSA components and therefore the vacuum limit becomes ambiguous. The inclusion of the subleading second tensor structure $F_4$ reduces the
	difference between the properties of the spatial and longitudinal projection of the vector mesons drastically. Due to numerical reasons only the spatial projection of the a$_1$ meson can be calculated. The mass of the spatial projection of the $a_1$ meson remains perfectly constant until the end of the light-quark coexistence region.
	
	We wish to point out that previous works in the DSE/BSE framework also find meson properties at finite chemical potential which satisfy the Silver-Blaze 
	property approximately \cite{Bender:1997jf,Maris:1997eg,Liu:2005cd,Jiang:2008rb,Jiang:2008zzd}. These works use effective interactions and a number of
	further approximations for the bound-state calculation. Nevertheless, qualitatively they deliver similar results than our approach at least up to the
	coexistence region. 
	
	It should be noted that the constant behaviour of masses and decay constants in the Silver-Blaze region up to a baryon chemical potential 
	of the mass of the nucleon minus its binding energy in nuclear matter is a highly non-trivial matter that relies on subtle cancellations between the
	chemical-potential dependence of the quarks, their interaction inside the mesons and the Bethe-Salpeter amplitudes (to be discussed below), which 
	together conspire to produce constant masses and decay constants. To our mind, it is very satisfying to find that this property holds in the functional 
	approach.

	\subsection{\label{sec:bsa}Bethe-Salpeter amplitudes}
	
	We discussed the chemical-potential dependency of the $\pi$ and $\sigma$ meson BSAs in detail already in Ref.~\cite{Gunkel:2019xnh}. Here we 
	reconsider these briefly and discuss updates due to our improved truncation scheme and detail in addition the corresponding behaviour of the 
	BSAs of the strange (pseudo-)scalars and the (axial-)vector mesons included in this work. A general property of all BS amplitudes for non-vanishing
	chemical potential is that they develop an imaginary part and they loose their symmetry properties under charge conjugation \cite{Gunkel:2019xnh}.
	
	In Fig.~\ref{fig:pion_bsa_finite_mu}, we display the real part of 
	all three medium-BSA components of the $\pi$ meson for fixed $P^2=-m_\pi^2$ plotted against the relative momentum $p$ between the quarks for 
	different chemical potentials. For a given chemical potential we furthermore plot amplitudes with different angles $\hat{p}\hat{P}$ between 
	total and relative momentum. The spread of the different line types is therefore a direct measure for the angular dependence of the amplitudes.   
	All three components show a similar behavior: With increasing chemical potential all become larger in the infrared, they all spread more in 
	the mid-momentum region and do not react to chemical potential in the ultraviolet momentum region. The strength of the infrared increase, however, 
	is different for the different BSA components. While $G^\textup{s}$ almost doubles its magnitude and is therefore comparable in strength to the 
	leading BSA component $E$ at large chemical potential, $I$ increases only weakly. Together this underlines the importance of taking $G^\textup{s}$
	into account, especially at large chemical potential. For small chemical potentials all BSA components connect smoothly to the vacuum limit.
	
	The $K$ meson BSAs (not shown) behave qualitatively similar as the $\pi$ BSA's but the changes are much less pronounced. Most importantly, 
	$G^\textup{s}$ remains weak and does not become comparable to the $E$ component. For the $\bar{K}$ meson $G^\textup{s}$ even decreases whereas $I$ shows a stronger increase compared to the $K$ meson. The $I$- and $G^\textup{s}$- BSA components of the $\sigma$ 
	meson decrease marginally in the infrared and increase their spread in the mid-momentum region for increasing chemical potentials. 
	All other mesons BSA's and in particular those for the (axial-)vector mesons show a very weak chemical-potential 
	dependence in their real part, but a significant dependence in their imaginary part. This can be seen explicitly for the lading tensor 
	component of the $\rho$ meson in the lower panel of Fig.~\ref{fig:pion_bsa_finite_mu}. Thus, it is the imaginary part of the amplitude
	that balances the variations in the quark propagator with respect to chemical potential and therefore provides for Silver-Blaze property.

	\section{\label{sec:summary}Summary and conclusions}

	In this work we have studied the masses and decay constants of light and strange (pseudo-)scalar and (axial-)vector mesons at finite chemical 
	potential up to and into the coexistence region of the first order chiral phase transition. To this end we employed a coupled system of 
	Dyson-Schwinger and Bethe-Salpeter equations for the unquenched gluon propagator and $N_f=2+1$ quark flavours in a truncation which has been 
	discussed and probed already elsewhere \cite{Gunkel:2019xnh}. 
	
	For all meson types we find constant values
	for masses and spatial decay constants at least up to a baryon chemical potential equal to the mass of the nucleon minus its binding energy 
	in nuclear matter. Thus the Silver-Blaze property of QCD \cite{Cohen:1991nk,Cohen:2004qp} is at work. Since all input ingredients into
	the Bethe-Salpeter equation describing these mesons do depend on chemical potential, it is a highly non-trivial matter that the meson's
	Bethe-Salpeter amplitudes adapt and conspire such that observable quantities remain unaffected. To our mind, it is very satisfying to find 
	that this property holds in the functional approach.

	
	\begin{acknowledgments}
		We thank Richard Williams and Philipp Isserstedt for valuable discussions and Philipp Isserstedt for a careful reading of the manuscript.
		This work has been supported by the Helmholtz Graduate School for Hadron and Ion Research
		(HGS-HIRe) for FAIR, the GSI Helmholtzzentrum f\"{u}r Schwerionenforschung, 
		and the BMBF under contract No.~05P18RGFCA. 
	\end{acknowledgments}
	
	\appendix*
	
	
	\bibliography{vector_mesons_finite_mu}

\begin{thebibliography}{37}%
\makeatletter
\providecommand \@ifxundefined [1]{%
 \@ifx{#1\undefined}
}%
\providecommand \@ifnum [1]{%
 \ifnum #1\expandafter \@firstoftwo
 \else \expandafter \@secondoftwo
 \fi
}%
\providecommand \@ifx [1]{%
 \ifx #1\expandafter \@firstoftwo
 \else \expandafter \@secondoftwo
 \fi
}%
\providecommand \natexlab [1]{#1}%
\providecommand \enquote  [1]{``#1''}%
\providecommand \bibnamefont  [1]{#1}%
\providecommand \bibfnamefont [1]{#1}%
\providecommand \citenamefont [1]{#1}%
\providecommand \href@noop [0]{\@secondoftwo}%
\providecommand \href [0]{\begingroup \@sanitize@url \@href}%
\providecommand \@href[1]{\@@startlink{#1}\@@href}%
\providecommand \@@href[1]{\endgroup#1\@@endlink}%
\providecommand \@sanitize@url [0]{\catcode `\\12\catcode `\$12\catcode
  `\&12\catcode `\#12\catcode `\^12\catcode `\_12\catcode `\%12\relax}%
\providecommand \@@startlink[1]{}%
\providecommand \@@endlink[0]{}%
\providecommand \url  [0]{\begingroup\@sanitize@url \@url }%
\providecommand \@url [1]{\endgroup\@href {#1}{\urlprefix }}%
\providecommand \urlprefix  [0]{URL }%
\providecommand \Eprint [0]{\href }%
\providecommand \doibase [0]{https://doi.org/}%
\providecommand \selectlanguage [0]{\@gobble}%
\providecommand \bibinfo  [0]{\@secondoftwo}%
\providecommand \bibfield  [0]{\@secondoftwo}%
\providecommand \translation [1]{[#1]}%
\providecommand \BibitemOpen [0]{}%
\providecommand \bibitemStop [0]{}%
\providecommand \bibitemNoStop [0]{.\EOS\space}%
\providecommand \EOS [0]{\spacefactor3000\relax}%
\providecommand \BibitemShut  [1]{\csname bibitem#1\endcsname}%
\let\auto@bib@innerbib\@empty
\bibitem [{\citenamefont {Rapp}\ and\ \citenamefont
  {Wambach}(2000)}]{Rapp:1999ej}%
  \BibitemOpen
  \bibfield  {author} {\bibinfo {author} {\bibfnamefont {R.}~\bibnamefont
  {Rapp}}\ and\ \bibinfo {author} {\bibfnamefont {J.}~\bibnamefont {Wambach}},\
  }\href {https://doi.org/10.1007/0-306-47101-9_1} {\bibfield  {journal}
  {\bibinfo  {journal} {Adv. Nucl. Phys.}\ }\textbf {\bibinfo {volume} {25}},\
  \bibinfo {pages} {1} (\bibinfo {year} {2000})},\ \Eprint
  {https://arxiv.org/abs/hep-ph/9909229} {arXiv:hep-ph/9909229 [hep-ph]}
  \BibitemShut {NoStop}%
\bibitem [{\citenamefont {Leupold}\ \emph {et~al.}(2010)\citenamefont
  {Leupold}, \citenamefont {Metag},\ and\ \citenamefont
  {Mosel}}]{Leupold:2009kz}%
  \BibitemOpen
  \bibfield  {author} {\bibinfo {author} {\bibfnamefont {S.}~\bibnamefont
  {Leupold}}, \bibinfo {author} {\bibfnamefont {V.}~\bibnamefont {Metag}},\
  and\ \bibinfo {author} {\bibfnamefont {U.}~\bibnamefont {Mosel}},\ }\href
  {https://doi.org/10.1142/S0218301310014728} {\bibfield  {journal} {\bibinfo
  {journal} {Int. J. Mod. Phys. E}\ }\textbf {\bibinfo {volume} {19}},\
  \bibinfo {pages} {147} (\bibinfo {year} {2010})},\ \Eprint
  {https://arxiv.org/abs/0907.2388} {arXiv:0907.2388 [nucl-th]} \BibitemShut
  {NoStop}%
\bibitem [{\citenamefont {Friman}\ \emph {et~al.}(2011)\citenamefont {Friman},
  \citenamefont {Hohne}, \citenamefont {Knoll}, \citenamefont {Leupold},
  \citenamefont {Randrup}, \citenamefont {Rapp},\ and\ \citenamefont
  {Senger}}]{Friman:2011zz}%
  \BibitemOpen
  \bibinfo {editor} {\bibfnamefont {B.}~\bibnamefont {Friman}}, \bibinfo
  {editor} {\bibfnamefont {C.}~\bibnamefont {Hohne}}, \bibinfo {editor}
  {\bibfnamefont {J.}~\bibnamefont {Knoll}}, \bibinfo {editor} {\bibfnamefont
  {S.}~\bibnamefont {Leupold}}, \bibinfo {editor} {\bibfnamefont
  {J.}~\bibnamefont {Randrup}}, \bibinfo {editor} {\bibfnamefont
  {R.}~\bibnamefont {Rapp}},\ and\ \bibinfo {editor} {\bibfnamefont
  {P.}~\bibnamefont {Senger}},\ eds.,\ \href
  {https://doi.org/10.1007/978-3-642-13293-3} {\emph {\bibinfo {title} {{The
  CBM physics book: Compressed baryonic matter in laboratory experiments}}}},\
  Vol.\ \bibinfo {volume} {814}\ (\bibinfo {year} {2011})\BibitemShut {NoStop}%
\bibitem [{\citenamefont {Jung}\ \emph {et~al.}(2017)\citenamefont {Jung},
  \citenamefont {Rennecke}, \citenamefont {Tripolt}, \citenamefont {von
  Smekal},\ and\ \citenamefont {Wambach}}]{Jung:2016yxl}%
  \BibitemOpen
  \bibfield  {author} {\bibinfo {author} {\bibfnamefont {C.}~\bibnamefont
  {Jung}}, \bibinfo {author} {\bibfnamefont {F.}~\bibnamefont {Rennecke}},
  \bibinfo {author} {\bibfnamefont {R.-A.}\ \bibnamefont {Tripolt}}, \bibinfo
  {author} {\bibfnamefont {L.}~\bibnamefont {von Smekal}},\ and\ \bibinfo
  {author} {\bibfnamefont {J.}~\bibnamefont {Wambach}},\ }\href
  {https://doi.org/10.1103/PhysRevD.95.036020} {\bibfield  {journal} {\bibinfo
  {journal} {Phys. Rev. D}\ }\textbf {\bibinfo {volume} {95}},\ \bibinfo
  {pages} {036020} (\bibinfo {year} {2017})},\ \Eprint
  {https://arxiv.org/abs/1610.08754} {arXiv:1610.08754 [hep-ph]} \BibitemShut
  {NoStop}%
\bibitem [{\citenamefont {Jung}\ and\ \citenamefont {von
  Smekal}(2019)}]{Jung:2019nnr}%
  \BibitemOpen
  \bibfield  {author} {\bibinfo {author} {\bibfnamefont {C.}~\bibnamefont
  {Jung}}\ and\ \bibinfo {author} {\bibfnamefont {L.}~\bibnamefont {von
  Smekal}},\ }\href {https://doi.org/10.1103/PhysRevD.100.116009} {\bibfield
  {journal} {\bibinfo  {journal} {Phys. Rev. D}\ }\textbf {\bibinfo {volume}
  {100}},\ \bibinfo {pages} {116009} (\bibinfo {year} {2019})},\ \Eprint
  {https://arxiv.org/abs/1909.13712} {arXiv:1909.13712 [hep-ph]} \BibitemShut
  {NoStop}%
\bibitem [{\citenamefont {Gunkel}\ \emph {et~al.}(2019)\citenamefont {Gunkel},
  \citenamefont {Fischer},\ and\ \citenamefont {Isserstedt}}]{Gunkel:2019xnh}%
  \BibitemOpen
  \bibfield  {author} {\bibinfo {author} {\bibfnamefont {P.~J.}\ \bibnamefont
  {Gunkel}}, \bibinfo {author} {\bibfnamefont {C.~S.}\ \bibnamefont
  {Fischer}},\ and\ \bibinfo {author} {\bibfnamefont {P.}~\bibnamefont
  {Isserstedt}},\ }\href {https://doi.org/10.1140/epja/i2019-12868-1}
  {\bibfield  {journal} {\bibinfo  {journal} {Eur. Phys. J. A}\ }\textbf
  {\bibinfo {volume} {55}},\ \bibinfo {pages} {169} (\bibinfo {year} {2019})},\
  \Eprint {https://arxiv.org/abs/1907.08110} {arXiv:1907.08110 [hep-ph]}
  \BibitemShut {NoStop}%
\bibitem [{\citenamefont {Chen}\ \emph {et~al.}(2020)\citenamefont {Chen},
  \citenamefont {Qin},\ and\ \citenamefont {Liu}}]{Chen:2020afc}%
  \BibitemOpen
  \bibfield  {author} {\bibinfo {author} {\bibfnamefont {L.-f.}\ \bibnamefont
  {Chen}}, \bibinfo {author} {\bibfnamefont {S.-X.}\ \bibnamefont {Qin}},\ and\
  \bibinfo {author} {\bibfnamefont {Y.-x.}\ \bibnamefont {Liu}},\ }\href@noop
  {} {\  (\bibinfo {year} {2020})},\ \Eprint {https://arxiv.org/abs/2006.10582}
  {arXiv:2006.10582 [hep-ph]} \BibitemShut {NoStop}%
\bibitem [{\citenamefont {Cohen}\ \emph {et~al.}(1992)\citenamefont {Cohen},
  \citenamefont {Furnstahl},\ and\ \citenamefont {Griegel}}]{Cohen:1991nk}%
  \BibitemOpen
  \bibfield  {author} {\bibinfo {author} {\bibfnamefont {T.~D.}\ \bibnamefont
  {Cohen}}, \bibinfo {author} {\bibfnamefont {R.~J.}\ \bibnamefont
  {Furnstahl}},\ and\ \bibinfo {author} {\bibfnamefont {D.~K.}\ \bibnamefont
  {Griegel}},\ }\href {https://doi.org/10.1103/PhysRevC.45.1881} {\bibfield
  {journal} {\bibinfo  {journal} {Phys. Rev. C}\ }\textbf {\bibinfo {volume}
  {45}},\ \bibinfo {pages} {1881} (\bibinfo {year} {1992})}\BibitemShut
  {NoStop}%
\bibitem [{\citenamefont {Cohen}(2004)}]{Cohen:2004qp}%
  \BibitemOpen
  \bibfield  {author} {\bibinfo {author} {\bibfnamefont {T.~D.}\ \bibnamefont
  {Cohen}},\ }\href@noop {} {}\bibinfo {howpublished}
  {\href{http://arxiv.org/abs/hep-ph/0405043}{arXiv:hep-ph/0405043}} (\bibinfo
  {year} {2004})\BibitemShut {NoStop}%
\bibitem [{\citenamefont {Fromm}\ \emph {et~al.}(2013)\citenamefont {Fromm},
  \citenamefont {Langelage}, \citenamefont {Lottini}, \citenamefont {Neuman},\
  and\ \citenamefont {Philipsen}}]{Fromm:2012eb}%
  \BibitemOpen
  \bibfield  {author} {\bibinfo {author} {\bibfnamefont {M.}~\bibnamefont
  {Fromm}}, \bibinfo {author} {\bibfnamefont {J.}~\bibnamefont {Langelage}},
  \bibinfo {author} {\bibfnamefont {S.}~\bibnamefont {Lottini}}, \bibinfo
  {author} {\bibfnamefont {M.}~\bibnamefont {Neuman}},\ and\ \bibinfo {author}
  {\bibfnamefont {O.}~\bibnamefont {Philipsen}},\ }\href
  {https://doi.org/10.1103/PhysRevLett.110.122001} {\bibfield  {journal}
  {\bibinfo  {journal} {Phys. Rev. Lett.}\ }\textbf {\bibinfo {volume} {110}},\
  \bibinfo {pages} {122001} (\bibinfo {year} {2013})},\ \Eprint
  {https://arxiv.org/abs/1207.3005} {arXiv:1207.3005 [hep-lat]} \BibitemShut
  {NoStop}%
\bibitem [{\citenamefont {Eichmann}\ \emph {et~al.}(2016)\citenamefont
  {Eichmann}, \citenamefont {Fischer},\ and\ \citenamefont
  {Welzbacher}}]{Eichmann:2015kfa}%
  \BibitemOpen
  \bibfield  {author} {\bibinfo {author} {\bibfnamefont {G.}~\bibnamefont
  {Eichmann}}, \bibinfo {author} {\bibfnamefont {C.~S.}\ \bibnamefont
  {Fischer}},\ and\ \bibinfo {author} {\bibfnamefont {C.~A.}\ \bibnamefont
  {Welzbacher}},\ }\href {https://doi.org/10.1103/PhysRevD.93.034013}
  {\bibfield  {journal} {\bibinfo  {journal} {Phys. Rev. D}\ }\textbf {\bibinfo
  {volume} {93}},\ \bibinfo {pages} {034013} (\bibinfo {year} {2016})},\
  \Eprint {https://arxiv.org/abs/1509.02082} {arXiv:1509.02082 [hep-ph]}
  \BibitemShut {NoStop}%
\bibitem [{\citenamefont {Fischer}(2019)}]{Fischer:2018sdj}%
  \BibitemOpen
  \bibfield  {author} {\bibinfo {author} {\bibfnamefont {C.~S.}\ \bibnamefont
  {Fischer}},\ }\href {https://doi.org/10.1016/j.ppnp.2019.01.002} {\bibfield
  {journal} {\bibinfo  {journal} {Prog. Part. Nucl. Phys.}\ }\textbf {\bibinfo
  {volume} {105}},\ \bibinfo {pages} {1} (\bibinfo {year} {2019})},\ \Eprint
  {https://arxiv.org/abs/1810.12938} {arXiv:1810.12938 [hep-ph]} \BibitemShut
  {NoStop}%
\bibitem [{\citenamefont {Fischer}(2009)}]{Fischer:2009wc}%
  \BibitemOpen
  \bibfield  {author} {\bibinfo {author} {\bibfnamefont {C.~S.}\ \bibnamefont
  {Fischer}},\ }\href {https://doi.org/10.1103/PhysRevLett.103.052003}
  {\bibfield  {journal} {\bibinfo  {journal} {Phys. Rev. Lett.}\ }\textbf
  {\bibinfo {volume} {103}},\ \bibinfo {pages} {052003} (\bibinfo {year}
  {2009})},\ \Eprint {https://arxiv.org/abs/0904.2700} {arXiv:0904.2700
  [hep-ph]} \BibitemShut {NoStop}%
\bibitem [{\citenamefont {Fischer}\ \emph {et~al.}(2010)\citenamefont
  {Fischer}, \citenamefont {Maas},\ and\ \citenamefont
  {Mueller}}]{Fischer:2010fx}%
  \BibitemOpen
  \bibfield  {author} {\bibinfo {author} {\bibfnamefont {C.~S.}\ \bibnamefont
  {Fischer}}, \bibinfo {author} {\bibfnamefont {A.}~\bibnamefont {Maas}},\ and\
  \bibinfo {author} {\bibfnamefont {J.~A.}\ \bibnamefont {Mueller}},\ }\href
  {https://doi.org/10.1140/epjc/s10052-010-1343-1} {\bibfield  {journal}
  {\bibinfo  {journal} {Eur. Phys. J. C}\ }\textbf {\bibinfo {volume} {68}},\
  \bibinfo {pages} {165} (\bibinfo {year} {2010})},\ \Eprint
  {https://arxiv.org/abs/1003.1960} {arXiv:1003.1960 [hep-ph]} \BibitemShut
  {NoStop}%
\bibitem [{\citenamefont {Fischer}\ and\ \citenamefont
  {Mueller}(2011)}]{Fischer:2011pk}%
  \BibitemOpen
  \bibfield  {author} {\bibinfo {author} {\bibfnamefont {C.~S.}\ \bibnamefont
  {Fischer}}\ and\ \bibinfo {author} {\bibfnamefont {J.~A.}\ \bibnamefont
  {Mueller}},\ }\href {https://doi.org/10.1103/PhysRevD.84.054013} {\bibfield
  {journal} {\bibinfo  {journal} {Phys. Rev. D}\ }\textbf {\bibinfo {volume}
  {84}},\ \bibinfo {pages} {054013} (\bibinfo {year} {2011})},\ \Eprint
  {https://arxiv.org/abs/1106.2700} {arXiv:1106.2700 [hep-ph]} \BibitemShut
  {NoStop}%
\bibitem [{\citenamefont {Fischer}\ \emph {et~al.}(2011)\citenamefont
  {Fischer}, \citenamefont {Luecker},\ and\ \citenamefont
  {Mueller}}]{Fischer:2011mz}%
  \BibitemOpen
  \bibfield  {author} {\bibinfo {author} {\bibfnamefont {C.~S.}\ \bibnamefont
  {Fischer}}, \bibinfo {author} {\bibfnamefont {J.}~\bibnamefont {Luecker}},\
  and\ \bibinfo {author} {\bibfnamefont {J.~A.}\ \bibnamefont {Mueller}},\
  }\href {https://doi.org/10.1016/j.physletb.2011.07.039} {\bibfield  {journal}
  {\bibinfo  {journal} {Phys. Lett. B}\ }\textbf {\bibinfo {volume} {702}},\
  \bibinfo {pages} {438} (\bibinfo {year} {2011})},\ \Eprint
  {https://arxiv.org/abs/1104.1564} {arXiv:1104.1564 [hep-ph]} \BibitemShut
  {NoStop}%
\bibitem [{\citenamefont {Fischer}\ and\ \citenamefont
  {Luecker}(2013)}]{Fischer:2012vc}%
  \BibitemOpen
  \bibfield  {author} {\bibinfo {author} {\bibfnamefont {C.~S.}\ \bibnamefont
  {Fischer}}\ and\ \bibinfo {author} {\bibfnamefont {J.}~\bibnamefont
  {Luecker}},\ }\href {https://doi.org/10.1016/j.physletb.2012.11.054}
  {\bibfield  {journal} {\bibinfo  {journal} {Phys. Lett. B}\ }\textbf
  {\bibinfo {volume} {718}},\ \bibinfo {pages} {1036} (\bibinfo {year}
  {2013})},\ \Eprint {https://arxiv.org/abs/1206.5191} {arXiv:1206.5191
  [hep-ph]} \BibitemShut {NoStop}%
\bibitem [{\citenamefont {Fischer}\ \emph {et~al.}(2014)\citenamefont
  {Fischer}, \citenamefont {Luecker},\ and\ \citenamefont
  {Welzbacher}}]{Fischer:2014ata}%
  \BibitemOpen
  \bibfield  {author} {\bibinfo {author} {\bibfnamefont {C.~S.}\ \bibnamefont
  {Fischer}}, \bibinfo {author} {\bibfnamefont {J.}~\bibnamefont {Luecker}},\
  and\ \bibinfo {author} {\bibfnamefont {C.~A.}\ \bibnamefont {Welzbacher}},\
  }\href {https://doi.org/10.1103/PhysRevD.90.034022} {\bibfield  {journal}
  {\bibinfo  {journal} {Phys. Rev. D}\ }\textbf {\bibinfo {volume} {90}},\
  \bibinfo {pages} {034022} (\bibinfo {year} {2014})},\ \Eprint
  {https://arxiv.org/abs/1405.4762} {arXiv:1405.4762 [hep-ph]} \BibitemShut
  {NoStop}%
\bibitem [{\citenamefont {M\"uller}\ \emph {et~al.}(2013)\citenamefont
  {M\"uller}, \citenamefont {Buballa},\ and\ \citenamefont
  {Wambach}}]{Muller:2013pya}%
  \BibitemOpen
  \bibfield  {author} {\bibinfo {author} {\bibfnamefont {D.}~\bibnamefont
  {M\"uller}}, \bibinfo {author} {\bibfnamefont {M.}~\bibnamefont {Buballa}},\
  and\ \bibinfo {author} {\bibfnamefont {J.}~\bibnamefont {Wambach}},\ }\href
  {https://doi.org/10.1140/epja/i2013-13096-5} {\bibfield  {journal} {\bibinfo
  {journal} {Eur. Phys. J. A}\ }\textbf {\bibinfo {volume} {49}},\ \bibinfo
  {pages} {96} (\bibinfo {year} {2013})},\ \Eprint
  {https://arxiv.org/abs/1303.2693} {arXiv:1303.2693 [hep-ph]} \BibitemShut
  {NoStop}%
\bibitem [{\citenamefont {M\"uller}\ \emph {et~al.}(2016)\citenamefont
  {M\"uller}, \citenamefont {Buballa},\ and\ \citenamefont
  {Wambach}}]{Muller:2016fdr}%
  \BibitemOpen
  \bibfield  {author} {\bibinfo {author} {\bibfnamefont {D.}~\bibnamefont
  {M\"uller}}, \bibinfo {author} {\bibfnamefont {M.}~\bibnamefont {Buballa}},\
  and\ \bibinfo {author} {\bibfnamefont {J.}~\bibnamefont {Wambach}},\
  }\href@noop {} {}\bibinfo {howpublished}
  {\href{http://arxiv.org/abs/1603.02865}{arXiv:1603.02865 [hep-ph]}} (\bibinfo
  {year} {2016})\BibitemShut {NoStop}%
\bibitem [{\citenamefont {Maris}\ \emph
  {et~al.}(1998{\natexlab{a}})\citenamefont {Maris}, \citenamefont {Roberts},\
  and\ \citenamefont {Tandy}}]{Maris:1997hd}%
  \BibitemOpen
  \bibfield  {author} {\bibinfo {author} {\bibfnamefont {P.}~\bibnamefont
  {Maris}}, \bibinfo {author} {\bibfnamefont {C.~D.}\ \bibnamefont {Roberts}},\
  and\ \bibinfo {author} {\bibfnamefont {P.~C.}\ \bibnamefont {Tandy}},\ }\href
  {https://doi.org/10.1016/S0370-2693(97)01535-9} {\bibfield  {journal}
  {\bibinfo  {journal} {Phys. Lett. B}\ }\textbf {\bibinfo {volume} {420}},\
  \bibinfo {pages} {267} (\bibinfo {year} {1998}{\natexlab{a}})},\ \Eprint
  {https://arxiv.org/abs/nucl-th/9707003} {arXiv:nucl-th/9707003 [nucl-th]}
  \BibitemShut {NoStop}%
\bibitem [{\citenamefont {Llewellyn~Smith}(1969)}]{Smith:1969}%
  \BibitemOpen
  \bibfield  {author} {\bibinfo {author} {\bibfnamefont {C.}~\bibnamefont
  {Llewellyn~Smith}},\ }\href@noop {} {\bibfield  {journal} {\bibinfo
  {journal} {Annals Phys.}\ }\textbf {\bibinfo {volume} {53}},\ \bibinfo
  {pages} {521} (\bibinfo {year} {1969})}\BibitemShut {NoStop}%
\bibitem [{\citenamefont {Fischer}\ \emph {et~al.}(2005)\citenamefont
  {Fischer}, \citenamefont {Watson},\ and\ \citenamefont
  {Cassing}}]{Fischer:2005en}%
  \BibitemOpen
  \bibfield  {author} {\bibinfo {author} {\bibfnamefont {C.~S.}\ \bibnamefont
  {Fischer}}, \bibinfo {author} {\bibfnamefont {P.}~\bibnamefont {Watson}},\
  and\ \bibinfo {author} {\bibfnamefont {W.}~\bibnamefont {Cassing}},\ }\href
  {https://doi.org/10.1103/PhysRevD.72.094025} {\bibfield  {journal} {\bibinfo
  {journal} {Phys. Rev. D}\ }\textbf {\bibinfo {volume} {72}},\ \bibinfo
  {pages} {094025} (\bibinfo {year} {2005})},\ \Eprint
  {https://arxiv.org/abs/hep-ph/0509213} {arXiv:hep-ph/0509213 [hep-ph]}
  \BibitemShut {NoStop}%
\bibitem [{\citenamefont {Williams}(2010)}]{Williams:2009wx}%
  \BibitemOpen
  \bibfield  {author} {\bibinfo {author} {\bibfnamefont {R.}~\bibnamefont
  {Williams}},\ }\href {https://doi.org/10.1051/epjconf/20100303005} {\bibfield
   {journal} {\bibinfo  {journal} {EPJ Web Conf.}\ }\textbf {\bibinfo {volume}
  {3}},\ \bibinfo {pages} {03005} (\bibinfo {year} {2010})},\ \Eprint
  {https://arxiv.org/abs/0912.3494} {arXiv:0912.3494 [hep-ph]} \BibitemShut
  {NoStop}%
\bibitem [{\citenamefont {Maris}\ and\ \citenamefont
  {Tandy}(2001)}]{Maris:2001rq}%
  \BibitemOpen
  \bibfield  {author} {\bibinfo {author} {\bibfnamefont {P.}~\bibnamefont
  {Maris}}\ and\ \bibinfo {author} {\bibfnamefont {P.~C.}\ \bibnamefont
  {Tandy}},\ }in\ \href@noop {} {\emph {\bibinfo {booktitle} {{Research Program
  at the Erwin Schrödinger Institute on Confinement Vienna, Austria, May
  5-July 17, 2000}}}}\ (\bibinfo {year} {2001})\ \Eprint
  {https://arxiv.org/abs/nucl-th/0109035} {arXiv:nucl-th/0109035 [nucl-th]}
  \BibitemShut {NoStop}%
\bibitem [{\citenamefont {Maris}\ \emph
  {et~al.}(1998{\natexlab{b}})\citenamefont {Maris}, \citenamefont {Roberts},\
  and\ \citenamefont {Schmidt}}]{Maris:1997eg}%
  \BibitemOpen
  \bibfield  {author} {\bibinfo {author} {\bibfnamefont {P.}~\bibnamefont
  {Maris}}, \bibinfo {author} {\bibfnamefont {C.~D.}\ \bibnamefont {Roberts}},\
  and\ \bibinfo {author} {\bibfnamefont {S.~M.}\ \bibnamefont {Schmidt}},\
  }\href {https://doi.org/10.1103/PhysRevC.57.R2821} {\bibfield  {journal}
  {\bibinfo  {journal} {Phys. Rev. C}\ }\textbf {\bibinfo {volume} {57}},\
  \bibinfo {pages} {R2821} (\bibinfo {year} {1998}{\natexlab{b}})},\ \Eprint
  {https://arxiv.org/abs/nucl-th/9801059} {arXiv:nucl-th/9801059 [nucl-th]}
  \BibitemShut {NoStop}%
\bibitem [{\citenamefont {Maris}\ and\ \citenamefont
  {Tandy}(1999)}]{Maris:1999nt}%
  \BibitemOpen
  \bibfield  {author} {\bibinfo {author} {\bibfnamefont {P.}~\bibnamefont
  {Maris}}\ and\ \bibinfo {author} {\bibfnamefont {P.~C.}\ \bibnamefont
  {Tandy}},\ }\href {https://doi.org/10.1103/PhysRevC.60.055214} {\bibfield
  {journal} {\bibinfo  {journal} {Phys. Rev. C}\ }\textbf {\bibinfo {volume}
  {60}},\ \bibinfo {pages} {055214} (\bibinfo {year} {1999})},\ \Eprint
  {https://arxiv.org/abs/nucl-th/9905056} {arXiv:nucl-th/9905056 [nucl-th]}
  \BibitemShut {NoStop}%
\bibitem [{\citenamefont {Nakanishi}(1965)}]{Nakanishi:1965zza}%
  \BibitemOpen
  \bibfield  {author} {\bibinfo {author} {\bibfnamefont {N.}~\bibnamefont
  {Nakanishi}},\ }\href {https://doi.org/10.1103/PhysRev.138.B1182} {\bibfield
  {journal} {\bibinfo  {journal} {Phys. Rev.}\ }\textbf {\bibinfo {volume}
  {138}},\ \bibinfo {pages} {B1182} (\bibinfo {year} {1965})}\BibitemShut
  {NoStop}%
\bibitem [{\citenamefont {Son}\ and\ \citenamefont
  {Stephanov}(2002{\natexlab{a}})}]{Son:2001ff}%
  \BibitemOpen
  \bibfield  {author} {\bibinfo {author} {\bibfnamefont {D.~T.}\ \bibnamefont
  {Son}}\ and\ \bibinfo {author} {\bibfnamefont {M.~A.}\ \bibnamefont
  {Stephanov}},\ }\href {https://doi.org/10.1103/PhysRevLett.88.202302}
  {\bibfield  {journal} {\bibinfo  {journal} {Phys. Rev. Lett.}\ }\textbf
  {\bibinfo {volume} {88}},\ \bibinfo {pages} {202302} (\bibinfo {year}
  {2002}{\natexlab{a}})},\ \Eprint {https://arxiv.org/abs/0111100}
  {arXiv:0111100 [hep-ph]} \BibitemShut {NoStop}%
\bibitem [{\citenamefont {Son}\ and\ \citenamefont
  {Stephanov}(2002{\natexlab{b}})}]{Son:2002ci}%
  \BibitemOpen
  \bibfield  {author} {\bibinfo {author} {\bibfnamefont {D.~T.}\ \bibnamefont
  {Son}}\ and\ \bibinfo {author} {\bibfnamefont {M.~A.}\ \bibnamefont
  {Stephanov}},\ }\href {https://doi.org/10.1103/PhysRevD.66.076011} {\bibfield
   {journal} {\bibinfo  {journal} {Phys. Rev. D}\ }\textbf {\bibinfo {volume}
  {66}},\ \bibinfo {pages} {076011} (\bibinfo {year} {2002}{\natexlab{b}})},\
  \Eprint {https://arxiv.org/abs/0204226} {arXiv:0204226 [hep-ph]} \BibitemShut
  {NoStop}%
\bibitem [{\citenamefont {Fischer}\ \emph {et~al.}(2009)\citenamefont
  {Fischer}, \citenamefont {Nickel},\ and\ \citenamefont
  {Williams}}]{Fischer:2008sp}%
  \BibitemOpen
  \bibfield  {author} {\bibinfo {author} {\bibfnamefont {C.~S.}\ \bibnamefont
  {Fischer}}, \bibinfo {author} {\bibfnamefont {D.}~\bibnamefont {Nickel}},\
  and\ \bibinfo {author} {\bibfnamefont {R.}~\bibnamefont {Williams}},\ }\href
  {https://doi.org/10.1140/epjc/s10052-008-0821-1} {\bibfield  {journal}
  {\bibinfo  {journal} {Eur. Phys. J. C}\ }\textbf {\bibinfo {volume} {60}},\
  \bibinfo {pages} {47} (\bibinfo {year} {2009})},\ \Eprint
  {https://arxiv.org/abs/0807.3486} {arXiv:0807.3486 [hep-ph]} \BibitemShut
  {NoStop}%
\bibitem [{\citenamefont {Otto}\ \emph {et~al.}(2020)\citenamefont {Otto},
  \citenamefont {Oertel},\ and\ \citenamefont {Schaefer}}]{Otto:2019zjy}%
  \BibitemOpen
  \bibfield  {author} {\bibinfo {author} {\bibfnamefont {K.}~\bibnamefont
  {Otto}}, \bibinfo {author} {\bibfnamefont {M.}~\bibnamefont {Oertel}},\ and\
  \bibinfo {author} {\bibfnamefont {B.-J.}\ \bibnamefont {Schaefer}},\ }\href
  {https://doi.org/10.1103/PhysRevD.101.103021} {\bibfield  {journal} {\bibinfo
   {journal} {Phys. Rev. D}\ }\textbf {\bibinfo {volume} {101}},\ \bibinfo
  {pages} {103021} (\bibinfo {year} {2020})},\ \Eprint
  {https://arxiv.org/abs/1910.11929} {arXiv:1910.11929 [hep-ph]} \BibitemShut
  {NoStop}%
\bibitem [{\citenamefont {Contant}\ and\ \citenamefont
  {Huber}(2020)}]{Contant:2019lwf}%
  \BibitemOpen
  \bibfield  {author} {\bibinfo {author} {\bibfnamefont {R.}~\bibnamefont
  {Contant}}\ and\ \bibinfo {author} {\bibfnamefont {M.~Q.}\ \bibnamefont
  {Huber}},\ }\href {https://doi.org/10.1103/PhysRevD.101.014016} {\bibfield
  {journal} {\bibinfo  {journal} {Phys. Rev. D}\ }\textbf {\bibinfo {volume}
  {101}},\ \bibinfo {pages} {014016} (\bibinfo {year} {2020})},\ \Eprint
  {https://arxiv.org/abs/1909.12796} {arXiv:1909.12796 [hep-ph]} \BibitemShut
  {NoStop}%
\bibitem [{\citenamefont {Bender}\ \emph {et~al.}(1998)\citenamefont {Bender},
  \citenamefont {Poulis}, \citenamefont {Roberts}, \citenamefont {Schmidt},\
  and\ \citenamefont {Thomas}}]{Bender:1997jf}%
  \BibitemOpen
  \bibfield  {author} {\bibinfo {author} {\bibfnamefont {A.}~\bibnamefont
  {Bender}}, \bibinfo {author} {\bibfnamefont {G.~I.}\ \bibnamefont {Poulis}},
  \bibinfo {author} {\bibfnamefont {C.~D.}\ \bibnamefont {Roberts}}, \bibinfo
  {author} {\bibfnamefont {S.~M.}\ \bibnamefont {Schmidt}},\ and\ \bibinfo
  {author} {\bibfnamefont {A.~W.}\ \bibnamefont {Thomas}},\ }\href
  {https://doi.org/10.1016/S0370-2693(98)00546-2} {\bibfield  {journal}
  {\bibinfo  {journal} {Phys. Lett. B}\ }\textbf {\bibinfo {volume} {431}},\
  \bibinfo {pages} {263} (\bibinfo {year} {1998})},\ \Eprint
  {https://arxiv.org/abs/nucl-th/9710069} {arXiv:nucl-th/9710069 [nucl-th]}
  \BibitemShut {NoStop}%
\bibitem [{\citenamefont {Liu}\ \emph {et~al.}(2005)\citenamefont {Liu},
  \citenamefont {Chao}, \citenamefont {Chang},\ and\ \citenamefont
  {Yuan}}]{Liu:2005cd}%
  \BibitemOpen
  \bibfield  {author} {\bibinfo {author} {\bibfnamefont {Y.-X.}\ \bibnamefont
  {Liu}}, \bibinfo {author} {\bibfnamefont {J.-Y.}\ \bibnamefont {Chao}},
  \bibinfo {author} {\bibfnamefont {L.}~\bibnamefont {Chang}},\ and\ \bibinfo
  {author} {\bibfnamefont {W.}~\bibnamefont {Yuan}},\ }\href
  {https://doi.org/10.1088/0256-307X/22/1/014} {\bibfield  {journal} {\bibinfo
  {journal} {Chin. Phys. Lett.}\ }\textbf {\bibinfo {volume} {22}},\ \bibinfo
  {pages} {46} (\bibinfo {year} {2005})}\BibitemShut {NoStop}%
\bibitem [{\citenamefont {Jiang}\ \emph
  {et~al.}(2008{\natexlab{a}})\citenamefont {Jiang}, \citenamefont {Shi},
  \citenamefont {Li}, \citenamefont {Sun},\ and\ \citenamefont
  {Zong}}]{Jiang:2008rb}%
  \BibitemOpen
  \bibfield  {author} {\bibinfo {author} {\bibfnamefont {Y.}~\bibnamefont
  {Jiang}}, \bibinfo {author} {\bibfnamefont {Y.-m.}\ \bibnamefont {Shi}},
  \bibinfo {author} {\bibfnamefont {H.}~\bibnamefont {Li}}, \bibinfo {author}
  {\bibfnamefont {W.-m.}\ \bibnamefont {Sun}},\ and\ \bibinfo {author}
  {\bibfnamefont {H.-s.}\ \bibnamefont {Zong}},\ }\href
  {https://doi.org/10.1103/PhysRevD.78.116005} {\bibfield  {journal} {\bibinfo
  {journal} {Phys. Rev. D}\ }\textbf {\bibinfo {volume} {78}},\ \bibinfo
  {pages} {116005} (\bibinfo {year} {2008}{\natexlab{a}})},\ \Eprint
  {https://arxiv.org/abs/0810.0750} {arXiv:0810.0750 [nucl-th]} \BibitemShut
  {NoStop}%
\bibitem [{\citenamefont {Jiang}\ \emph
  {et~al.}(2008{\natexlab{b}})\citenamefont {Jiang}, \citenamefont {Shi},
  \citenamefont {Feng}, \citenamefont {Sun},\ and\ \citenamefont
  {Zong}}]{Jiang:2008zzd}%
  \BibitemOpen
  \bibfield  {author} {\bibinfo {author} {\bibfnamefont {Y.}~\bibnamefont
  {Jiang}}, \bibinfo {author} {\bibfnamefont {Y.-m.}\ \bibnamefont {Shi}},
  \bibinfo {author} {\bibfnamefont {H.-t.}\ \bibnamefont {Feng}}, \bibinfo
  {author} {\bibfnamefont {W.-m.}\ \bibnamefont {Sun}},\ and\ \bibinfo {author}
  {\bibfnamefont {H.-s.}\ \bibnamefont {Zong}},\ }\href
  {https://doi.org/10.1103/PhysRevC.78.025214} {\bibfield  {journal} {\bibinfo
  {journal} {Phys. Rev. C}\ }\textbf {\bibinfo {volume} {78}},\ \bibinfo
  {pages} {025214} (\bibinfo {year} {2008}{\natexlab{b}})}\BibitemShut
  {NoStop}%
\end{thebibliography}%
	
\end{document}